# Nb₂O₅ high-k dielectric enabled electric field engineering of β-Ga₂O₃ metal-insulator-semiconductor (MIS) diode

Prabhans Tiwari,[1] Jayeeta Biswas,[1] Chandan Joishi,[1,2] and Saurabh Lodha[1, a)]

[1)] *Department of Electrical Engineering, Indian Institute of Technology Bombay, Mumbai, Maharashtra 400076, India*

[2)] *Department of Electrical and Computer Engineering, The Ohio State University, Columbus, OH 43210, U.S.A.*

[a)] electronic mail: slodha@ee.iitb.ac.in

## ABSTRACT

We demonstrate an Nb₂O₅/β-Ga₂O₃ metal-insulator-semiconductor (MIS) hetero-junction diode with Nb₂O₅ as the high-k dielectric insulator for more efficient electric field management resulting in enhanced breakdown characteristics compared to a β-Ga₂O₃ Schottky barrier diode. The Nb₂O₅ dielectric films were grown using atomic layer deposition and exhibited a high dielectric constant of 50. The high dielectric constant resulted in a 5× lower electric field at the metal/dielectric interface in the MIS diode compared to the metal/ β-Ga₂O₃ interface in the Schottky barrier diode. With good electron conduction in forward bias enabled by the negative conduction band offset of Nb₂O₅ w.r.t β-Ga₂O₃, the MIS design led to a 3× improvement in the reverse blocking voltage with a slight trade-off in the specific on-resistance. Overall, a 3.3× increase in the power figure of merit was observed (3.25 MW/cm² for the Schottky diode and 10.8 MW/cm² for the MIS diode). A detailed analysis of the energy band line-up, and the forward and reverse current transport mechanisms are also presented using analytical modeling and 2-D TCAD simulations.

# I.    INTRODUCTION

The efficacy of a semiconductor material for high power electronics is generally evaluated by the well-known Baliga's figure of merits (FoMs)[1] for low and high switching frequencies, which are functions of the critical electric field ($E_C$). For a semiconductor with good transport properties, a large $E_C$ enables high FoMs and is therefore, a primary prerequisite for realizing efficient next generation power electronic devices. This understanding has led to a huge interest in beta-gallium oxide ($\beta$-$Ga_2O_3$), which is an ultra-wide band gap material ($E_g \sim 4.8$ eV) with a high theoretical $E_C$ of 8 MV/cm, bulk electron mobility of 250-350 $cm^2V^{-1}$ s$^{-1}$, and saturation velocity of 1-2$\times 10^7$ cm/s.[2–4] Besides, the material is also the only wide band gap semiconductor that can be grown from the melt similar to silicon, thereby facilitating high quality large area native substrates at low-cost for homoepitaxy.[5,6]

Power devices with vertical as well as lateral transport such as Schottky barrier diodes (SBDs)[7–9] and field-effect transistors (FETs)[10–12] have been demonstrated using $\beta$-$Ga_2O_3$ with excellent FoMs. The $E_C$ in these devices, however, is limited to 3-4 MV/cm. This is because the theoretical $E_C$ for $\beta$-$Ga_2O_3$, as predicted from the band-gap vs breakdown field characteristics is based on band-to-band tunneling across a rectifying p-n junction. A near flat valence band profile and absence of shallow acceptors hold back the realization of p-type $\beta$-$Ga_2O_3$. The lack of p-type doping, therefore, limits the $\beta$-$Ga_2O_3$ niche to uni-polar Schottky contacts. Since conventional Schottky barrier heights on $\beta$-$Ga_2O_3$ range from 1.1-1.5 eV ($<< E_g$), the corresponding $E_C$ is limited to 3-4 MV/cm beyond which significant leakage current flows across the metal-semiconductor interface. To overcome these limitations and realize the full breakdown strength of the material, metal-oxidized Schottky contacts to increase the Schottky barrier height[13], p-n hetero-junctions[14], and high-k dielectric-based metal-insulator-semiconductor (MIS) diodes[15] are being extensively studied.

The maximum contact barrier heights for the metal-oxidized Schottky contacts are intrinsically limited to ~2.4 eV by Fermi level pinning near mid-gap of $\beta$-$Ga_2O_3$.[13] For p-n hetero-junctions, the p-type materials explored till date have a lower $E_g$ and poor-interface w.r.t $\beta$-$Ga_2O_3$ that limits the breakdown characteristics. The high-k dielectric hetero-junctions, on the other hand, rely on the idea of continuity of electric displacement field over a high-k/$\beta$-$Ga_2O_3$ interface. Based on dielectric constant discontinuity, a low electric field can result at the metal/dielectric interface under reverse bias while maintaining a comparatively high field in the $\beta$-$Ga_2O_3$ channel. In the forward bias, a low/negative conduction band-offset of the high-k dielectric w.r.t. $\beta$-$Ga_2O_3$ enables reasonable current transport properties. The idea has been implemented to realize $E_C > 5.5$ MV/cm in $\beta$-$Ga_2O_3$ [15] and AlGaN based devices.[16] Also, this concept has been used in the design of high-k oxide field plated Schottky diodes.[17]

In this work, we utilize $Nb_2O_5$ as the high-k dielectric ($\epsilon_{Nb_2O_5} \sim 50$)[18] to demonstrate a Ni/$Nb_2O_5$/$\beta$-$Ga_2O_3$ MIS diode wherein the high dielectric constant enables 5$\times$ lower electric field ($\epsilon_{Nb_2O_5}/\epsilon_{Ga_2O_3}$) at the Schottky gate/$Nb_2O_5$ interface for a relatively high field in $\beta$-$Ga_2O_3$ (see Section I in supplementary for dielectric constant extraction). This efficient field management resulted in 3$\times$ improvement of the reverse blocking voltage with good forward conduction enabled by the nearly similar electron affinities of $Nb_2O_5$ and $\beta$-$Ga_2O_3$.

## II.     EXPERIMENTAL METHODS

For a comparative study, two 5x5 mm$^2$ Sn-doped β-Ga$_2$O$_3$ ($\bar{2}$01) wafers with a doping density of 2×10$^{17}$cm$^{-3}$ from Tamura corporation were used for fabrication of the Schottky and the MIS diode. First, the samples were organically cleaned by a 3-minute dip in methanol and acetone, followed by a dip in piranha solution (deionized (DI) water:30% H$_2$O$_2$:96% H$_2$SO$_4$ in 1:1:4 ratio) for 5 minutes. A 40 nm thick Nb$_2$O$_5$ film was deposited on one of the samples by thermal atomic layer deposition (ALD) using (tert-butylimido) tris (diethylamido) niobium (TBTDEN) and water at 200 ℃. The TBTDEN precursor was heated to 90 ℃ and an argon gas boost was used for efficient precursor delivery to the reaction chamber. The precursor was pulsed for 1 s, followed by a wait time of 20 s. Finally, water was pulsed for 0.06 s, followed by a purge time of 10 s. Post deposition, rapid thermal annealing was performed at 650 ℃ for 1 minute in N$_2$ ambient. The dielectric properties were characterized using ellipsometry, X-ray photo-electron spectroscopy (XPS), grazing-incidence X-ray diffraction (GIXRD), and transmission electron microscopy (TEM) studies on the as-deposited and the annealed Nb$_2$O$_5$ films. The samples were then patterned using optical lithography to define Ni/Au gate contacts, followed by Ti/Au deposition to form the backside substrate contact. To achieve good ohmic contacts, the samples were annealed at 470 ℃ for 1 minute in N$_2$ ambient. Figs. 1(a) and 1(b) show the cross-section schematics of the fabricated devices.

## III.     RESULTS AND DISCUSSION

A thickness of 38 nm and refractive index of 2.36 (in good agreement with literature reports)[19] were confirmed for the Nb$_2$O$_5$ films by ellipsometry. The stoichiometry and film composition of Nb$_2$O$_5$ were studied by XPS measurements. Figs. 1(c) and 1(d) show the Nb 3d and O 1s XPS spectra of the annealed film. Nb 3d$_{5/2}$ and Nb 3d$_{3/2}$ peaks positioned at 206.87 eV and 209.62 eV, and O 1s peak at 529.90 eV confirmed the existence of Nb-O bonds.[20] Annealing Nb$_2$O$_5$ at 650 ℃ reduced the oxygen vacancy concentration in the film as seen through a comparison of XPS spectra of the as-deposited and the annealed films (see Section II in supplementary for XPS spectra of as-deposited film). This compensation of n-type conductivity led to lower diode leakage at reverse bias for the annealed MIS diodes compared to their non-annealed counterparts (see Fig. S2(c) in supplementary). To study the morphology of Nb$_2$O$_5$ on β-Ga$_2$O$_3$, GIXRD was performed using Cu Kα radiation in a Rigaku Smartlab Diffractometer and the patterns are shown in Figs 1(e) and 1(f). For the as-deposited film, no peaks were observed in the spectrum, indicating its amorphous nature. However, two distinct peaks were observed for the annealed film corresponding to the (001) and (200) planes of Nb$_2$O$_5$, indicating its poly-crystalline nature. The observed peaks are in good agreement with the reflections reported for the orthorhombic phase of Nb$_2$O$_5$.[21] The morphology of the films was confirmed through TEM images (see Section III in supplementary).

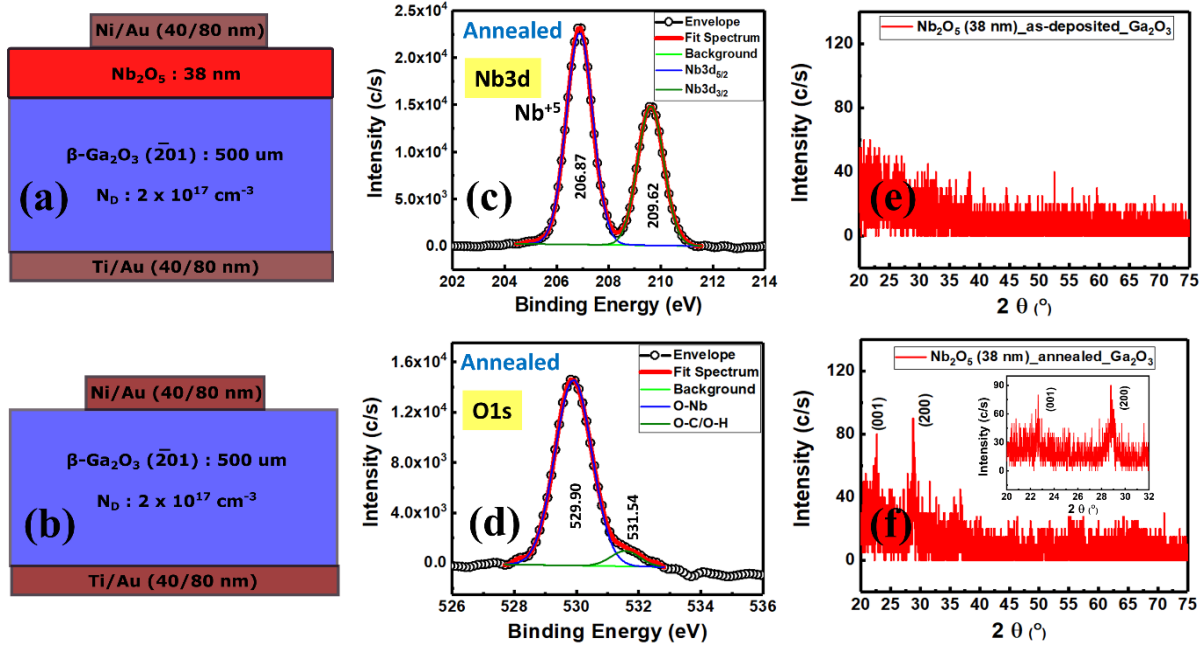

FIG. 1. Cross-section schematics of (a) the MIS diode and (b) the Schottky diode. (c) Nb 3d and (d) O 1s XPS spectra of 38 nm $Nb_2O_5$ deposited at 200 ℃ and annealed at 650 ℃. GIXRD pattern of (e) $Nb_2O_5$ deposited at 200 ℃, showing its amorphous nature and (f) $Nb_2O_5$ deposited at 200 ℃ and annealed at 650 ℃, showing its poly-crystalline nature. The inset shows a magnified view of the observed peaks.

Figs. 2(a) and 2(b) show the zero-bias energy band diagrams generated using a Synopsys TCAD simulator by fitting the forward bias current density-voltage ($J$-$V$) experimental data (see Section IV in supplementary for the simulation details). Barrier heights of 0.9 eV and 1.01 eV were extracted from simulations for the MIS and the Schottky diodes respectively. A relatively small negative conduction band offset of 0.12 eV was obtained for $Nb_2O_5$ w.r.t. β-$Ga_2O_3$ (difference between electron affinities of $Nb_2O_5$ and β-$Ga_2O_3$), which ensures good forward conduction. From the 2-D electric field profiles in Figs. 2(c) and 2(d), it can be seen that there is approximately a 5× reduction in the field at the metal/dielectric interface for the MIS diode compared to the metal/semiconductor interface of the Schottky diode, made possible by the high dielectric constant of $Nb_2O_5$. The reduced field at the interface ensures a wider tunneling barrier thereby resulting in reduced leakage current at similar voltages, thus allowing the device to breakdown at higher reverse voltages with a corresponding high electric field in the β-$Ga_2O_3$ channel.

The simulated reverse bias energy band diagrams ($V_R$ = -40 V) are plotted in Figs. 3(a) and 3(b). A wider tunneling barrier is observed for the MIS diode compared to the Schottky diode. This reduces the probability of electron tunneling through the barrier and thus allows the MIS diode to reach higher voltages before breakdown by gate leakage. Fig. 3(c) shows the room temperature reverse bias $J$-$V$ characteristics for both the devices in logarithmic scale measured using a Keysight B1500 semiconductor parameter analyzer. The Schottky diode broke-down at a reverse voltage of 65 V, while the MIS diode was able to sustain voltages till 200 V (voltage measurements higher than 200 V were limited by instrument

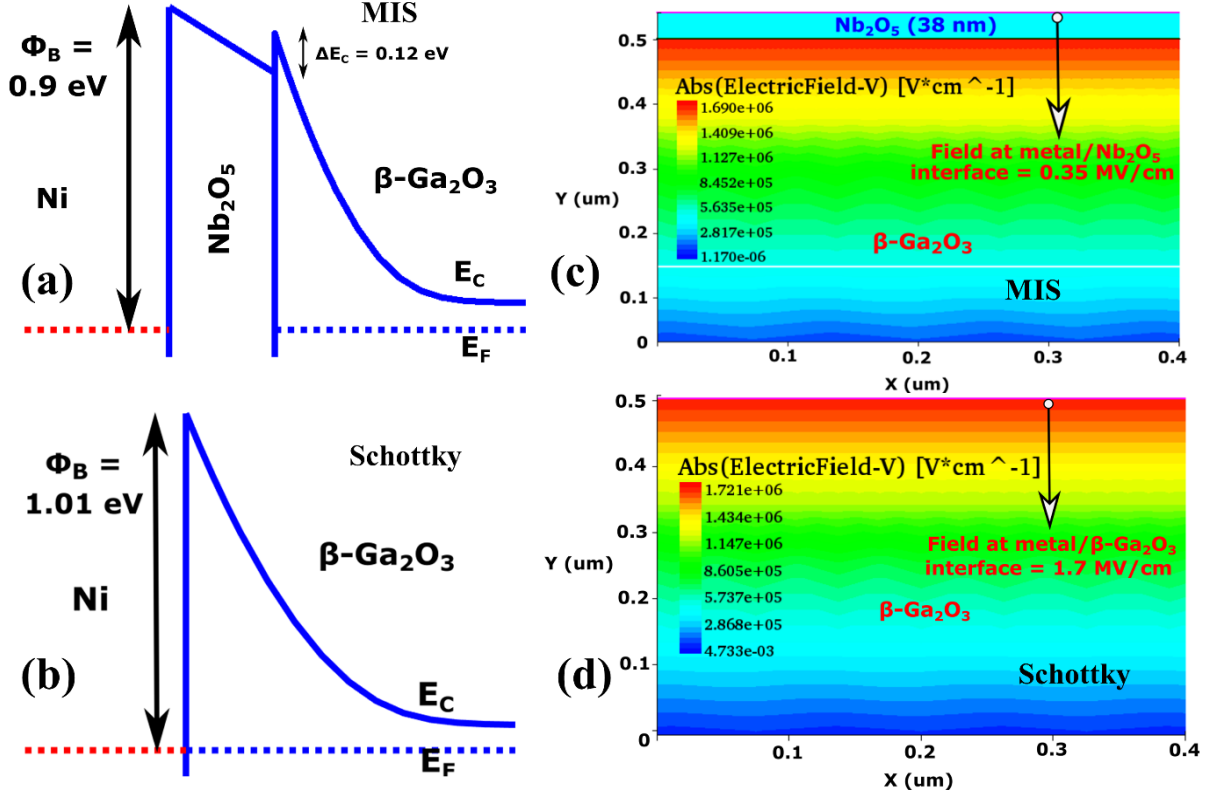

FIG. 2. Zero bias band diagram of (a) the MIS diode and (b) the Schottky diode. 2-D electric field profile under reverse bias of -40 V of (c) the MIS diode and (d) the Schottky diode

compliance). A numerical current leakage model based on WKB tunneling probability was used to fit the reverse leakage with barrier height as the fitting parameter.[22] The model showed excellent fit with the experimental data for barrier heights of 0.85 eV and 0.95 eV for the MIS and the Schottky diodes. Corresponding breakdown electric fields of 2.2 MV/cm and 3.8 MV/cm were obtained from 2-D TCAD simulations (see Section V in supplementary for the electric field simulation plots). Fig. 3(d) shows the room temperature forward *J-V* characteristics. Ideality factor of 1.09 and 1.07, $I_{ON}/I_{OFF}$ ratio of $10^{10}$ and $10^9$, and specific on-resistances ($R_{on}$) of 1.3 mΩ-cm$^2$ and 3.7 mΩ-cm$^2$ were measured for the Schottky and the MIS diodes respectively from the plot. No hysteresis was observed after performing dual sweep *J-V* measurements on the MIS diode, which indicated minimal trapping in the dielectric in the forward bias regime (see Section VI in supplementary for the plots). Power figures of merit ($V_{br}^2/R_{on}$) of 3.25 MW/cm$^2$ and 10.8 MW/cm$^2$ were obtained for the Schottky and the MIS diode respectively.

To understand the current transport mechanisms, temperature-dependent *J-V* measurements were carried out for both the devices between 25 ℃ and 160 ℃. The forward bias *J-V* characteristics were analyzed using a thermionic emission (*TE*) model[23]

$$J = J_S \left( e^{qV/nkT} - 1 \right) \quad (1)$$

where

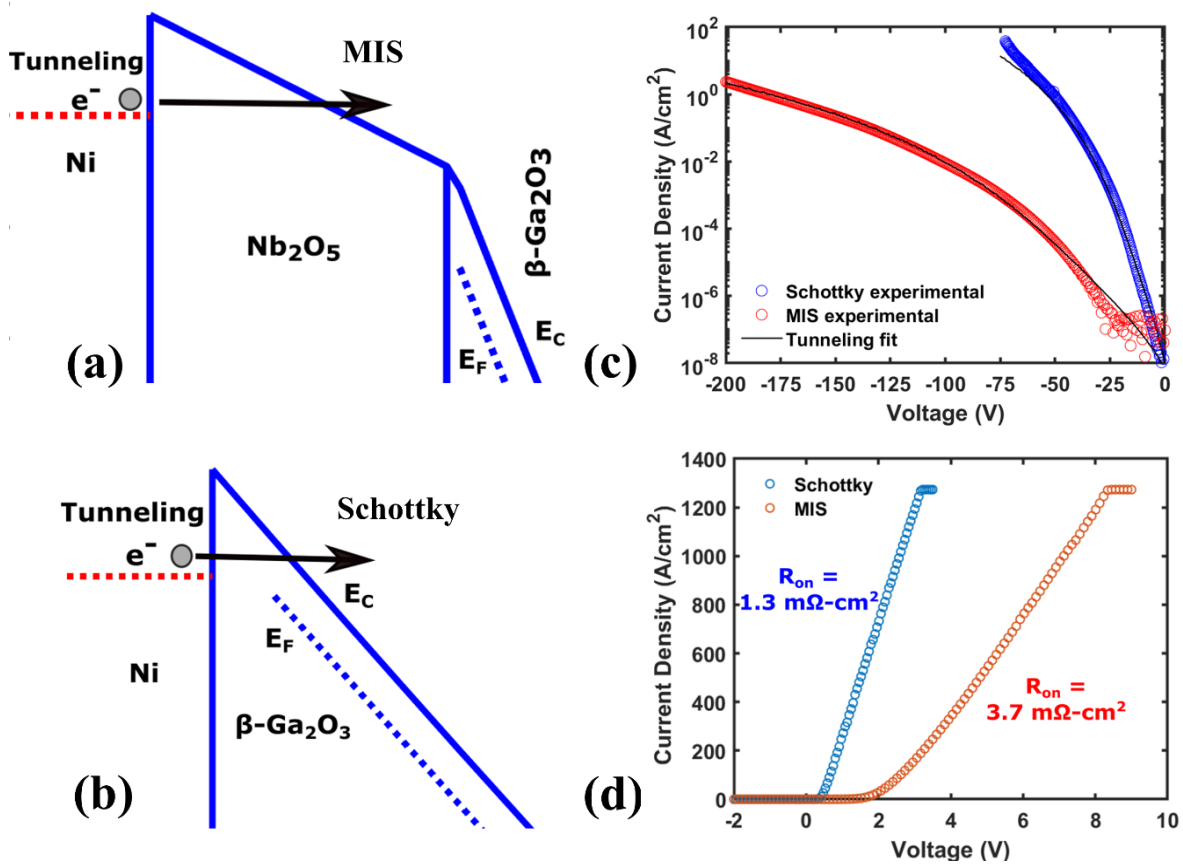

FIG. 3. (a) Reverse bias band diagram of the MIS diode showing the tunneling of electrons through a wider barrier under reverse bias ($V_R$ = -40 V), (b) Reverse bias band diagram of the Schottky diode showing the tunneling of electrons under reverse bias ($V_R$ = -40 V), (c) Room temperature $J$-$V$ characteristics under reverse bias and the corresponding tunneling fits, and (d) Room temperature $J$-$V$ characteristics under forward bias.

$$J_s = A^{**}T^2 e^{-q\Phi_B/kT} \qquad (2)$$

Here, $J_s$ is the reverse saturation current density, $A^{**}$ is the Richardson constant, $V$ is the applied bias, $n$ is the ideality factor, $T$ is the temperature, $k$ is the Boltzmann constant, $q$ is the electron charge, and $\Phi_B$ is the barrier height. To validate the $TE$ model, Eq. (2) is rearranged to construct the Richardson plot according to[23]

$$ln\left(J_s/T^2\right) = ln(A^{**}) - q\Phi_B/kT \qquad (3)$$

The Richardson constant and barrier height can be extracted from the intercept and slope of the linear fit to the Richardson plot. We implement a modified Richardson plot of $ln(J_s/T^2)$ versus $1/n(T)kT$ to account for the temperature dependence of $n$ that ensures a more accurate estimation of the Richardson constant.[24] The modified Richardson curve for the Schottky and the MIS diodes are plotted in Figs. 4(a) and 4(d). From a linear fit of the data points, Richardson constant of 39.7 A/cm$^2$/K$^2$ and a barrier height of 0.99 eV were extracted for the Schottky diode. The extracted Richardson constant is in good agreement with other experimental reports.[25] Similarly, Richardson constant of 6.5 A/cm$^2$/K$^2$ and a barrier height of 0.92 eV were extracted for the MIS diode, close to the calculated theoretical Richardson constant of 4.8 A/cm²/K² for Nb₂O₅ using an electron effective mass of 0.04 $m_o$.[26] Excellent fits were obtained for the experimental data using the $TE$ model as shown in Figs. 4(b) and 4(e) (see Section VII in supplementary for analysis of other conduction mechanisms) and the fitting parameters (barrier height and ideality factor) are shown in Figs. 4(c) and 4(f). The barrier height is constant across all temperatures for the Schottky diode, while the slight temperature dependence of the barrier height for the MIS diode could be due to barrier inhomogeneity.[27] The extracted barrier heights are in good agreement with the values obtained from TCAD simulation and reverse $J$-$V$ tunneling fits.(From TCAD simulation and reverse $J$-$V$ fits, barrier heights of 1.01 eV and 0.95 eV were obtained for the Schottky diode, while 0.9 eV and 0.85 eV were obtained for the MIS diode respectively.)

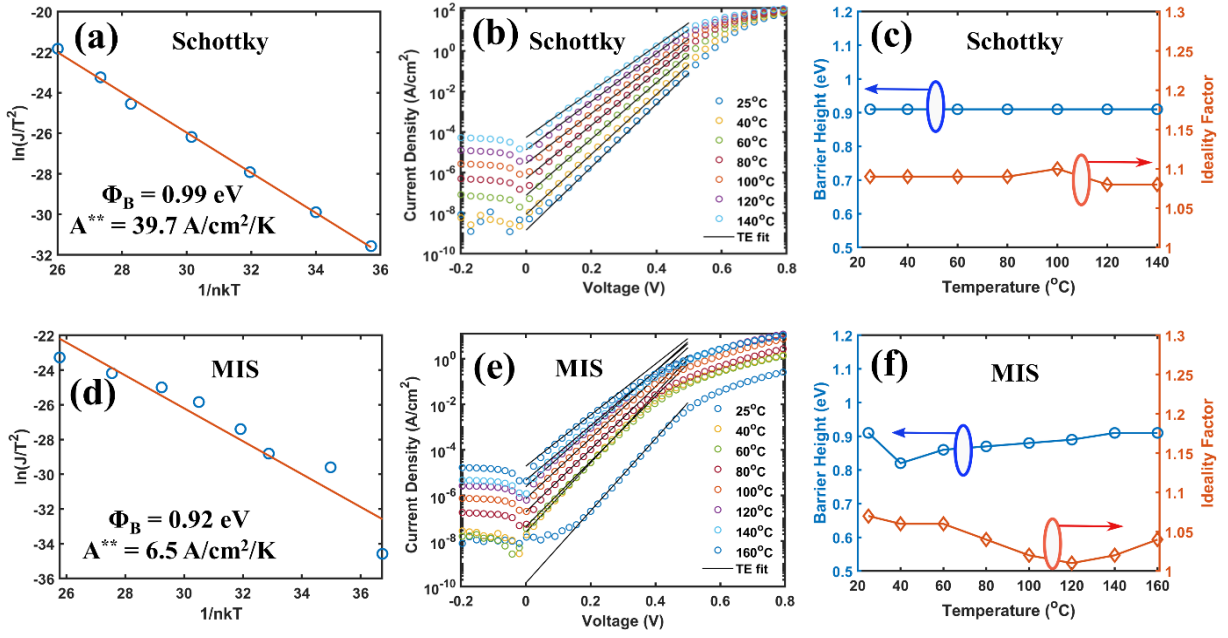

FIG. 4. (a) Modified Richardson plot of $ln(J_s/T^2)$ versus $1/n(T)kT$, (b) $TE$ fits for the temperature-dependent forward $J$-$V$ characteristics and (c) Extracted barrier height and ideality factor vs temperature from the $TE$ model fits of the Schottky diode. (d) Modified Richardson plot of $ln(J_s/T^2)$ versus $1/n(T)kT$, (e) $TE$ fits for the temperature-dependent forward $J$-$V$ characteristics and (f) Extracted barrier height and ideality factor vs temperature from the $TE$ model fits of the MIS diode.

The temperature-dependent reverse bias $J$-$V$ characteristics are plotted in Figs. 5(a), 5(b) and 5(c). For the Schottky diode, the numerical tunneling model as discussed in Fig. 3(c) was adequate to explain the $J$-$V$ characteristics for the entire temperature range. Barrier height of 0.96 eV was used to fit the experimental data. For the MIS diode, the room temperature $J$-$V$ characteristics showed excellent fit with the numerical tunneling model. However, between 40 ℃ and 80 ℃, Poole-Frenkel and hopping conduction models along with the tunneling model were needed to fit the data. Hopping conduction is due to the tunneling of electrons from one trap site to another in dielectric films. The hopping conduction mechanism is given by[28]

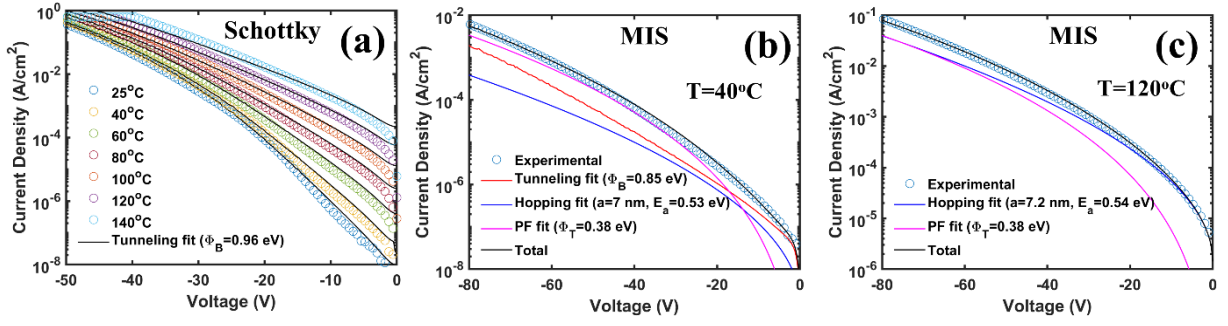

FIG. 5. Temperature-dependent reverse bias $J$-$V$ characteristics for (a) the Schottky diode between 25 ℃ and 140 ℃, (b) the MIS diode at 40 ℃, and (c) the MIS diode at 120 ℃ (see Section X in supplementary for the MIS fits at other temperatures). For the Schottky diode, the experimental data at all temperatures could be fitted with the numerical tunneling model. Poole-Frenkel and hopping conduction mechanisms, along with the tunneling model explain the characteristics between 40 ℃ and 80 ℃ for the MIS diode, while the characteristics above 80 ℃ can be explained by the Poole-Frenkel and hopping conduction models alone.

$$ J = qan_e v \, exp \left[ {qaE_{ox}}/{kT} - {E_a}/{kT} \right] \qquad (4) $$

where $a$ is the mean hopping distance (i.e., the mean spacing between trap sites), $n_e$ is the electron concentration in the conduction band of the dielectric, $v$ is the frequency of thermal vibrations of electrons at trap sites, $E_{ox}$ is the oxide field, $E_a$ is the activation energy and other notations are the same as defined before. A mean hoping distance ($a$) of 7 nm and $E_a = 0.53$ eV was used to fit the experimental data (see Section VIII in supplementary for parameter extraction). In Poole-Frenkel conduction mechanism, a trapped electron gets thermally excited into the dielectric conduction band under the influence of an electric field across the dielectric. The current density due to Poole-Frenkel emission is[28]

$$ J = CE \, exp \left[ {-q\left( \Phi_T - \beta_{PF}\sqrt{E} \right)}/{kT} \right] \qquad (5) $$

where $C$ is a constant, $E$ is the electric field in β-Ga$_2$O$_3$, $\Phi_T$ is the trap energy level and $\beta_{PF}$ is the Poole-Frenkel slope given by

$$ \beta_{PF} = \sqrt{{q}/{\pi \epsilon_i \epsilon_o}} \qquad (6) $$

where $\epsilon_i$ is the optical dielectric constant ((refractive index)$^2$), $\epsilon_o$ is the permittivity in vacuum and other notations are the same as defined before. A Poole-Frenkel slope of 4.2e-5 eVV$^{-1/2}$m$^{1/2}$ and trap barrier height of 0.38 eV was extracted from the experimental data (see Section IX in supplementary). The extracted slope is in good agreement with the theoretical slope of 4e-5 eVV$^{-1/2}$m$^{1/2}$ for β-Ga$_2$O$_3$.[29] The emergence of Poole-Frenkel and hopping conduction mechanisms above room temperature could be due to the higher energy requirements to overcome the trap barrier, and these mechanisms alone were sufficient to fit the characteristics above 80 ℃.

## IV. CONCLUSION

In conclusion, an $Nb_2O_5/\beta$-$Ga_2O_3$ hetero-junction diode for efficient electrostatic engineering of electric fields in the $\beta$-$Ga_2O_3$ drift layer was demonstrated to enhance reverse blocking characteristics. The MIS diode displayed a reverse blocking voltage 3× more than the Schottky diode with a slight increase in the specific on-resistance to 3.7 m$\Omega$-cm$^2$. Furthermore, a detailed understanding of the current transport mechanisms via temperature dependent forward and reverse *J-V* characteristics was also presented. The design/modeling strategy discussed here serves as a guide to design high-k based Schottky hetero-junction diodes wherein the $\beta$-$Ga_2O_3$ drift layer would exhibit near theoretical critical electric fields, necessary for next generation power devices.

## SUPPLEMENTARY MATERIAL

See supplementary material (below in the same file) for dielectric constant extraction, XPS and TEM analysis of $Nb_2O_5$ films, 2-D TCAD simulation details, breakdown electric field simulations, dual sweep forward bias *J-V* characteristics, parameter extraction for the hopping and Poole-Frenkel conduction mechanisms in the MIS diode in reverse bias and the temperature dependent reverse bias *J-V* fits for the MIS diode.

## ACKNOWLEDGMENTS

The authors acknowledge MeitY and DST, Government of India, for funding this work through the Nanoelectronics Network For Research and Application (NNetRA) project.

Supplementary Material

# Nb$_2$O$_5$ high-k dielectric enabled electric field engineering of β-Ga$_2$O$_3$ metal-insulator-semiconductor (MIS) diode

Prabhans Tiwari,[1] Jayeeta Biswas,[1] Chandan Joishi,[1,2] and Saurabh Lodha[1, a]

[3] *Department of Electrical Engineering, Indian Institute of Technology Bombay, Mumbai, Maharashtra 400076, India*

[4] *Department of Electrical and Computer Engineering, The Ohio State University, Columbus, OH 43210, U.S.A.*

[a] electronic mail: slodha@ee.iitb.ac.in

# I.   Extraction of the dielectric constant of Nb₂O₅

Capacitance-Voltage ($C$-$V$) measurements were carried out on Ni/Nb$_2$O$_5$ (as-deposited)/β-Ga$_2$O$_3$ and Ni/Nb$_2$O$_5$ (annealed)/β-Ga$_2$O$_3$ devices. Figs. S1(a) and S1(b) show the measured $C$-$V$ curves at 1 MHz. In order to compensate for the series resistance ($r_s$), the true capacitance $C$ was calculated from the measured capacitance $C_m$ using the following relation[1]

$$C_m = \frac{C}{(1+r_s G)^2 + (2\pi f r_s C)^2} \tag{S1}$$

From the maximum accumulation capacitance, dielectric constant values of (62.1 ± 0.6) and (53.2 ± 2.2) were extracted for as-deposited and annealed Nb$_2$O$_5$ respectively using the relation

$$k = \frac{C \times T_{ox}}{A \times \varepsilon_o} \tag{S2}$$

where, $k$ is the dielectric constant, $T_{ox}$ is the oxide thickness, $A$ is the device area and $\varepsilon_o$ is the permittivity of free space. The extracted values are in good agreement with the reported dielectric constant for Nb$_2$O$_5$.[2,3] The decrease in the dielectric constant after anneal can be attributed to the poly-crystalline nature of Nb$_2$O$_5$ film, which is in agreement with literature reports.[2,3]

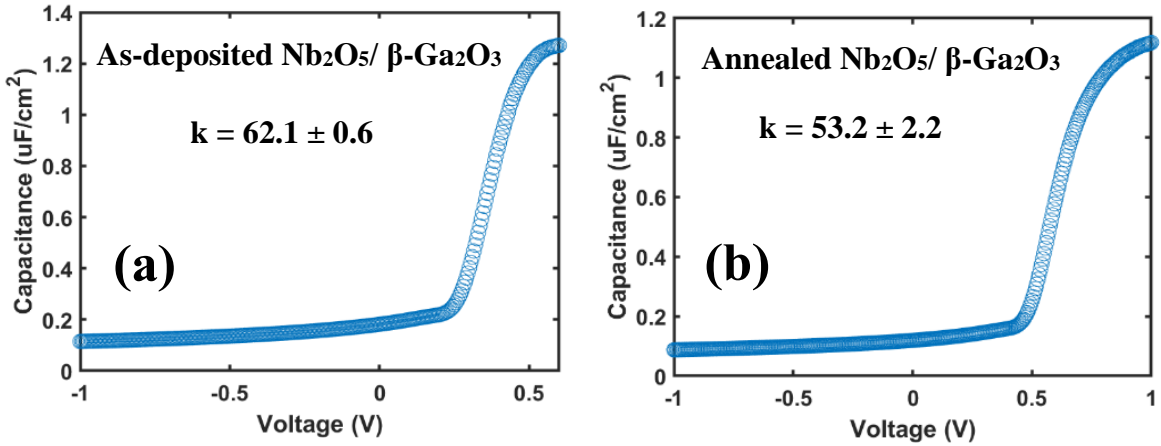

FIG. S1. $C$-$V$ characteristics of (a) Ni/Nb$_2$O$_5$ (as-deposited)/β-Ga$_2$O$_3$ and (b) Ni/Nb$_2$O$_5$ (annealed)/β-Ga$_2$O$_3$ devices at 1 MHz.

## II. XPS analysis of $Nb_2O_5$ film deposited on β-$Ga_2O_3$

Figs. S2(a) and S2(b) show the Nb 3d and O 1s XPS spectra of the as-deposited $Nb_2O_5$ film. Nb $3d_{5/2}$ and Nb $3d_{3/2}$ peak positions at 206.89 eV and 209.65 eV, and O 1s peak at 530.02 eV are similar to that reported in literature.[4] The as-deposited films were sub-stoichiometric with O/Nb ratio <2.5 consistent with literature reports.[4] The sub-stoichiometric nature of the as-deposited films can be attributed to oxygen vacancies,[5] reported to be positively charged and responsible for the unintentional n-type conductivity.[6,7] Post anneals (see the XPS plot in the main manuscript), an O/Nb ratio of 2.5 was observed, indicating stoichiometric film with reduced oxygen vacancy concentration. The reduction in oxygen vacancies for annealed $Nb_2O_5$ was consistent with the shift in the Nb 3d and O 1s spectral peaks towards lower binding energies, as well as a narrower full width at half maximum (FWHM) compared to the as-deposited one (see Table S1).[8] Thus, reduced leakage through $Nb_2O_5$ was observed for the annealed films due to compensation of n-type conductivity. For the diodes with the annealed films, ~20000× reduction in current density was observed at a reverse voltage of 50 V compared to the diodes with as-deposited films, as can be Fig. S2(c).

Table S1: XPS analysis of as-deposited and annealed $Nb_2O_5$ films.

| $Nb_2O_5$ film detail | Nb 3d | | | | O 1s | | | | Atomic concentration (%) | | | O/Nb ratio |
| | Nb $3d_{5/2}$ | | Nb $3d_{3/2}$ | | O/Nb bond | | O-C/O-H bond | | | | | |
| | Peak BE (eV) | FWHM (eV) | Peak BE (eV) | FWHM (eV) | Peak BE (eV) | FWHM (eV) | Peak BE (eV) | FWHM (eV) | Nb | O | C | |
| As-deposited | 206.89 | 1.25 | 209.65 | 1.22 | 530.02 | 1.38 | 531.62 | 1.14 | 25.29 | 60.24 | 14.48 | 2.38 |
| Annealed | 206.87 | 1.21 | 209.62 | 1.18 | 529.90 | 1.33 | 531.52 | 1.09 | 24.76 | 61.50 | 13.74 | 2.48 |

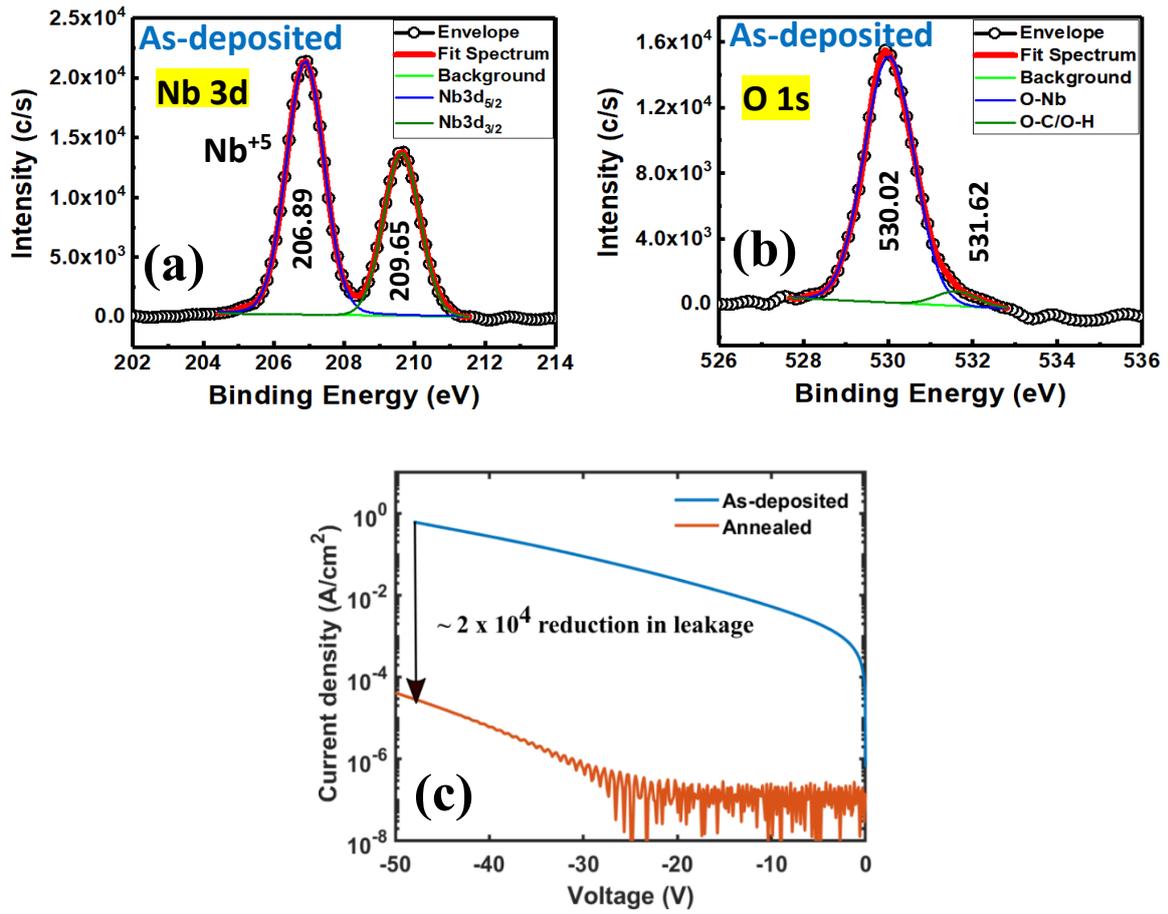

FIG. S2. (a) Nb 3d and (b) O 1s spectra of 38 nm Nb$_2$O$_5$ deposited on β-Ga$_2$O$_3$ at 200ºC. (c) Leakage current comparison of diodes with as-deposited and annealed Nb$_2$O$_5$ films. ~ 20000× reduction in leakage was observed for the annealed diodes at a reverse voltage of 50 V compared to the as-deposited ones.

# III.   GIXRD and TEM analysis of Nb₂O₅ films deposited on silicon

Figs. S3(a) and S3(b) show the GIXRD scans of as-deposited and annealed $Nb_2O_5$ films respectively. Absence of $Nb_2O_5$ peaks indicates the amorphous nature of the as-deposited film. However, three distinct peaks corresponding to the (001), (200) and (201) planes of orthorhombic $Nb_2O_5$ were observed for the films annealed at 650 °C, which indicates its poly-crystalline nature. The morphology was confirmed with TEM images, shown in Figs. S3(c)-S3(f). Lack of periodic atomic arrangement confirmed the amorphous nature of the as-deposited film, whereas multiple crystal planes were observed in the lattice images of the annealed film. From the images, (130), (200) and (201) planes were obtained from their corresponding d-spacing values of 5.2 Å, 3.1 Å and 2.4 Å respectively, which confirmed the poly-crystalline nature of the annealed film.

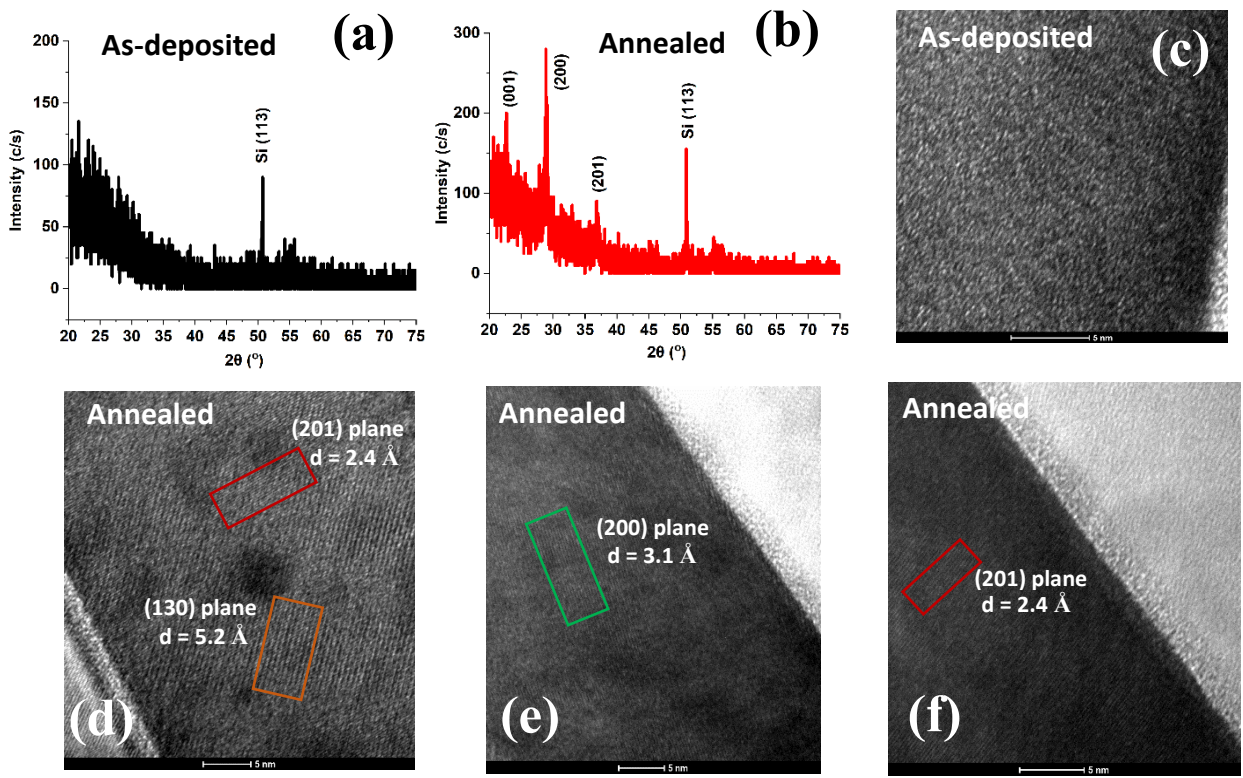

FIG. S3. GIXRD spectra of (a) $Nb_2O_5$ film deposited on silicon at 200 °C and (b) $Nb_2O_5$ film deposited on silicon at 200 °C and annealed at 650 °C. TEM images of (c) as-deposited and (d)-(f) annealed $Nb_2O_5$ films deposited on silicon.

# IV. 2-D TCAD simulation methodology

Thermionic emission and tunneling models were used in the 2-D TCAD simulations. The effective metal work function was first determined by simulating the Schottky diode. Using the same metal work function, the electron affinity (EA) of $Nb_2O_5$ (4.12 eV) was determined by comparing the MIS experimental data (in the exponential region) and the simulations for different electron affinities (3.9 eV – 4.3 eV is the reported range of EA values for $Nb_2O_5$[9]). The simulated forward bias curves and their fits with experimental data are shown in Figs. S4(a) and S4(b).

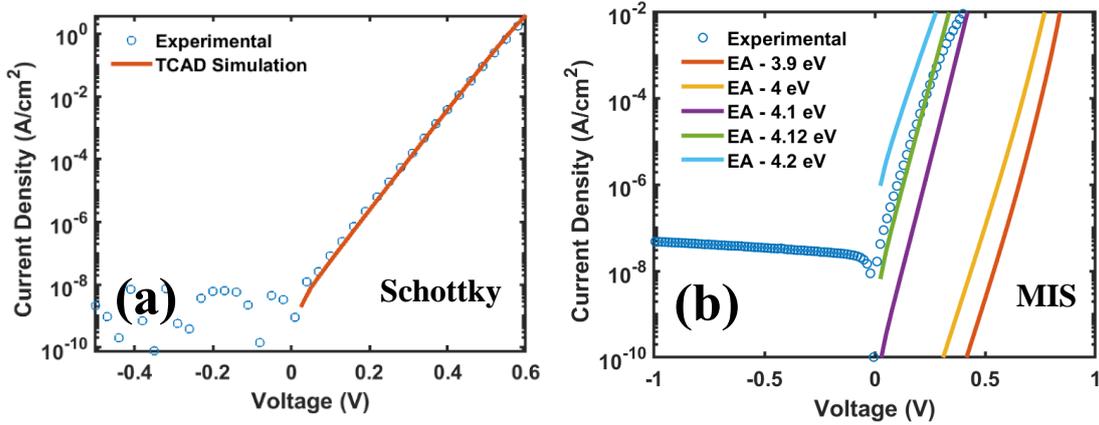

FIG. S4. (a) The simulated forward bias *J-V* curve fitted with the experimental data for the Schottky diode, and (b) the simulated forward bias *J-V* curves (in solid lines) for the MIS diode. Good fit with experimental data is obtained for an $Nb_2O_5$ electron affinity of 4.12 eV.

# V. Breakdown electric field simulation plots

Figs. S5(a) and S5(b) show the electric field simulations at breakdown voltage of -65 V and -200 V for the Schottky and the MIS diode respectively.

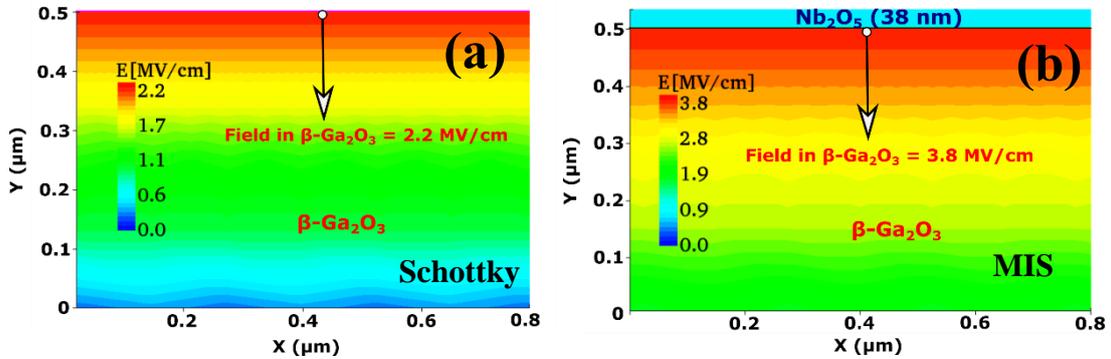

FIG. S5. Breakdown electric field simulation plots of (a) the Schottky diode, and (b) the MIS diode.

## VI. Dual sweep in the forward bias for the MIS diode

Dual sweep (forward and backward) was performed in the forward bias at two different sweep rates of 0.02 V/s and 0.5 V/s for the MIS diode as can be seen in Figs. S6(a) and S6(b). The forward and backward *J-V* characteristics overlap well, thus, confirming the non-existence of trapping and hysteresis in the diodes in the forward bias.

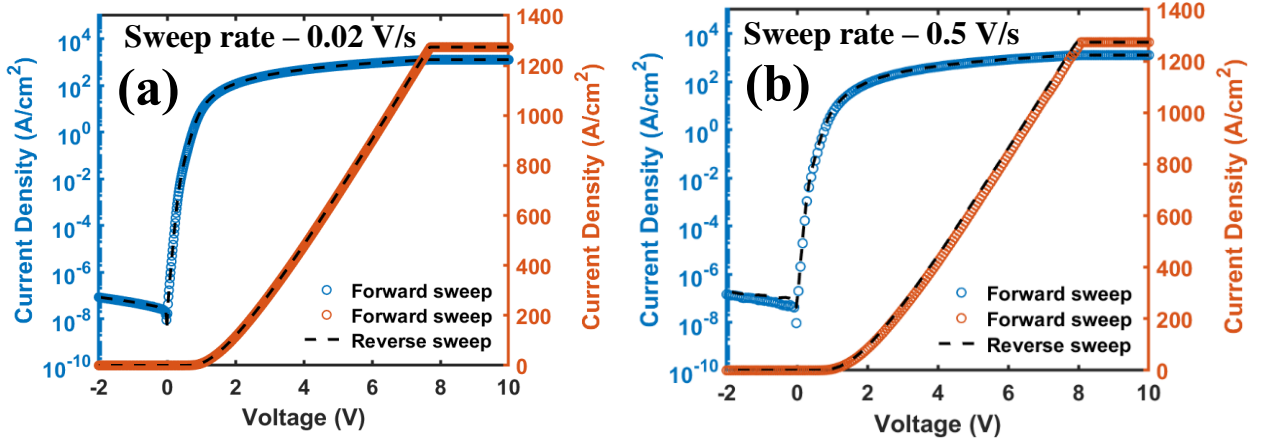

FIG. S6. Dual sweep forward bias *J-V* characteristics of the MIS diode at sweep rates of (a) 0.02 V/s and (b) 0.5 V/s.

## VII. Analysis of conduction mechanisms for the MIS diode in forward bias

Linear behavior in the log(*J*) vs *V* plot at low voltages usually indicates either thermionic emission (TE), thermionic field emission (TFE), field emission (FE) or defect assisted tunneling (DAT) conduction mechanisms. Also, we have analyzed space charge limited conduction mechanism and all the mechanisms are summarized in Table S2.

$E_{00}$ is the tunneling characteristic energy given by[11]

$$E_{00} = \frac{h}{4\pi}\sqrt{\frac{N}{m\epsilon}} \qquad (S3)$$

where *h* is Planck's constant, *N* is the semiconductor doping density, *m* is the tunneling mass and $\epsilon$ is the semiconductor dielectric constant.

$E_0$ is related to $E_{00}$ by the equation below[11]

$$E_0 = E_{00}\coth\left(\frac{E_{00}}{kT}\right) \qquad (S4)$$

Table S2: Overview of the possible conduction mechanisms in the MIS diode in forward bias

| Conduction mechanism | Equation | Extracted parameter | Dominance | Reason |
|---|---|---|---|---|
| TE[10] | $J = J_s\left(e^{qV/nkT} - 1\right);$ $J_s = AT^2\left(e^{-q\Phi_B/kT}\right)$ | 1. Richardson constant $A$ = 6.5 A/cm²/K 2. Barrier height $\Phi_B$ = 0.92 eV 3. $E_{00}/kT$ = 0.18 @ RT | Yes | 1. Good agreement with theoretical Richardson constant and barrier heights obtained from other methods 2. $E_{00}/kT < 1$ @ RT |
| FE[11] | $J = J_{FE(0)}\left(e^{qV/E_{00}}\right)$ | $E_{00}/kT$ = 0.18 @ RT | No | $E_{00}/kT < 1$ @ RT and strong temperature dependence of the curves |
| TFE[11] | $J = J_{TFE(0)}\left(e^{qV/E_0}\right)$ | 1. Barrier height $\Phi_B$ = 1.1 eV 2. $E_{00}/kT$ = 0.18 @ RT | No | $\Phi_B$ much higher than the values obtained from other methods and $E_{00}/kT < 1$ @ RT |
| DAT[12] | $J = J_{DAT(0)}\left(e^{qV/E_0}\right)$ | $E_{00}/kT$ = 0.18 @ RT | No | $E_{00}/kT < 1$ @ RT |
| SCLC[13] | $J_{Ohm} \propto V$ $J_{TFL} \propto V^2$ $J_{Child} \propto V^2$ | Three linear regions observed with slopes of 12, 6 and 1.5 | No | Log-log plot of the experimental data doesn't resemble the characteristics of SCLC conduction |

$E_{00}$ is related to the tunneling probability and the ratio $E_{00}/kT$ is a measure of the relative importance of the thermionic process in relation to the tunneling process. For FE and DAT conduction mechanisms, $E_{00}/kT \gg 1$, ~ 1 for TFE conduction mechanism, and $\ll 1$ for TE conduction mechanism.[12,14] Using doping density of $2\times10^{17}$ cm⁻³ of the substrate used in the experiment, the value of $E_{00}$ extracted was 4.72 meV and it is much less than the value of $kT$ at room temperature (26 meV). This fact combined with the strong temperature dependence of the exponential part of the *J-V* characteristics ruled out the FE conduction mechanism.

Using the values of $E_0$ extracted from the slope of the experimental data, good fit was obtained for the DAT mechanism as shown in Figure S7(a), however as $E_{00}/kT < 1$, this mechanism is unlikely to play a role in forward bias conduction.[12]

For SCLC, *J-V* characteristics plotted on a log-log plot should initially display a straight line with a slope of 1, followed by two regions of straight lines with a slope of 2.[13] However, we did not observe such behavior as can be seen from the log-log plot of the *J-V* characteristics of the MIS diodes as shown in Fig. S7(b), and thus we did not have any evidence to suggest the existence of space charge transport.

For TFE conduction mechanism, the saturation current density is given by the equation[11]

$$J_{TFE\,(0)} = \frac{AT\sqrt{\pi E_{00} q(\Phi_B - V - V_n)}}{k \cosh\left(\frac{E_{00}}{kT}\right)} \times \exp\left(-\frac{V_n}{kT} - \frac{\Phi_B - V_n}{E_0}\right) \qquad (S5)$$

where $A$ is the Richardson's constant, $\Phi_B$ is the barrier height, and $V_n$ is the energy of the Fermi level of the semiconductor measured with respect to the bottom of its conduction band. A plot of the logarithm of $J_{TFE\,(0)} \times \cosh(E_{00}/kT)/T$ vs. $1/E_0$ should be a straight line of slope ($\Phi_B - V_n$). For TE, the evidence of its existence would be evident from the modified Richardson plot as discussed in the manuscript. $R^2$ values of 0.92 and 0.90 were obtained for the linear fits in TFE plot and the modified Richardson plot respectively.

Barrier height of 1.1 eV was obtained from the linear fit in the TFE plot (shown in Fig. S7(c)) which is much higher than the barrier heights obtained from other methods (0.9 eV and 0.85 eV from the TCAD simulations and the reverse bias $J$-$V$ tunneling fits respectively). Also, since $E_{00}/kT <1$ at RT, it ruled out the possibility of TFE being the dominant conduction mechanism.

From the modified Richardson plot, Richardson constant of 6.5 A/cm$^2$/K was extracted which is in agreement with the theoretical Richardson constant of $Nb_2O_5$ and obtained barrier height of 0.92 eV was also in good agreement with the values obtained from other methods. The above findings along with the absence of other transport mechanisms suggested that TE should be the dominant conduction mechanism in the MIS diodes in forward bias.

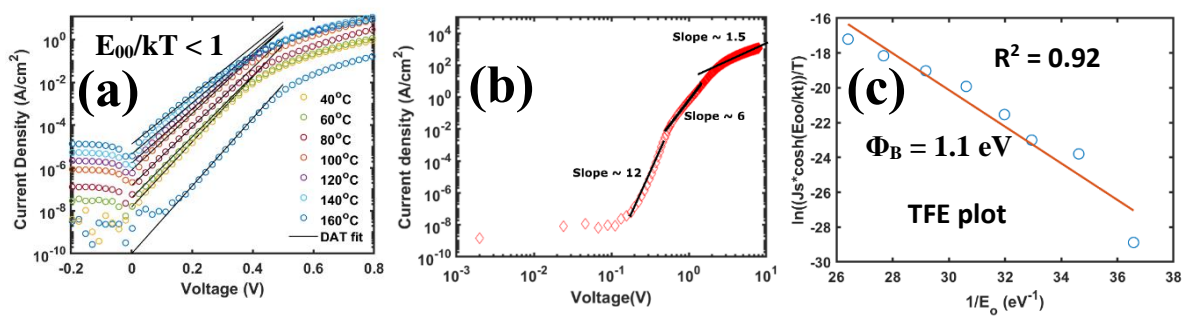

FIG. S7. (a) Forward bias $J$-$V$ fits for the MIS diode using DAT conduction mechanism, (b) log-log plot of the forward bias $J$-$V$ curve typically used to analyse SCLC conduction mechanism, and (b) TFE plot used for validation of TFE conduction mechanism.

# VIII. Parameter extraction for hopping conduction mechanism in the MIS diode under reverse bias

The hopping conduction mechanism is governed by the equation[13]

$$J = qan_e v \exp\left[\frac{qaE_{ox}}{kT} - \frac{E_a}{kT}\right] \qquad (S6)$$

where $a$ is the mean hopping distance, $n_e$ is the electron concentration in the conduction band of the dielectric, $v$ is the frequency of thermal vibration of electrons at trap sites, $E_{ox}$ is the electric field in the dielectric, $E_a$ is the activation energy, $k$ is the Boltzmann's constant, and $T$ is the temperature. As hopping conduction is due to the tunneling effect of trapped electrons hopping from one trap site to another in dielectric films, the electrons can transit even when the carrier energy is lower than the maximum energy of the potential barrier between two trapping sites. Excellent linear fits (shown by the black lines) were obtained for the characteristic plot of hopping conduction (ln($J$) vs $E_{ox}$) in low electric fields ($E_{ox}$<0.3 MV/cm), as shown in Fig. S8(a). From the slope of the plot, the mean hopping distance was extracted and the values were between 7 and 8 nm. From the Arrhenius plot in Fig. S8(b), an activation energy of 0.53 eV was extracted.

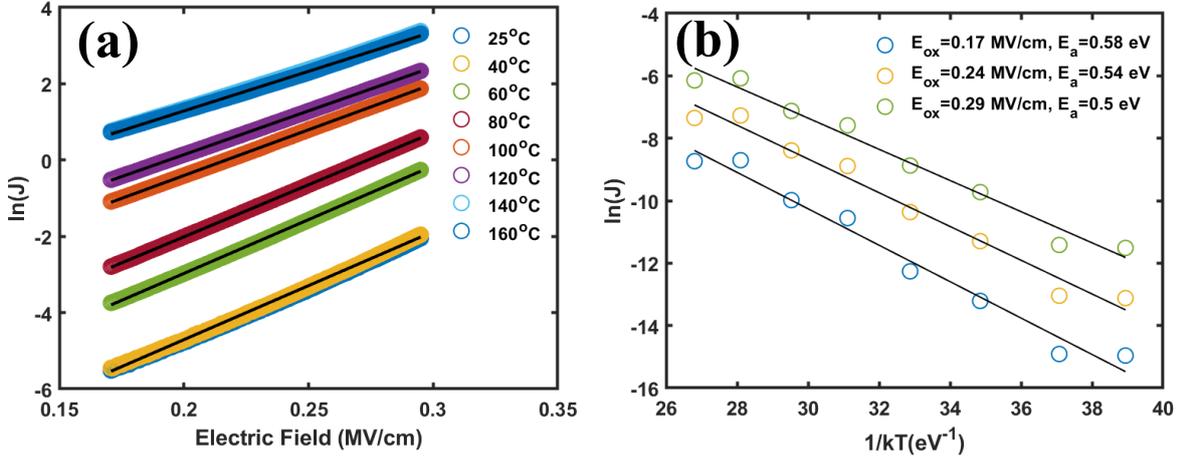

FIG. S8. (a) Characteristic plot of hopping conduction and (b) Arrhenius plot for the MIS diode under low reverse bias between 25 °C and 160 °C.

# IX. Parameter extraction for Poole-Frenkel conduction mechanism in the MIS diode under reverse bias

The current density due to Poole-Frenkel emission is[13]

$$J = CE \exp\left[\frac{-q(\Phi_T - \beta_{PF}\sqrt{E})}{kT}\right] \qquad (S7)$$

where $C$ is a constant, $E$ is the electric field, $\Phi_T$ is the trap energy level from the conduction band, and $\beta_{PF}$ is the Poole-Frenkel slope given by

$$\beta_{PF} = \sqrt{q/\pi\epsilon_i\epsilon_0} \qquad (S8)$$

Here, $q$ is the electron charge and $\epsilon_i$ is the optical dielectric constant.

As Poole-Frenkel conduction occurs due to the thermal excitation of trapped electrons into the conduction band of the dielectric, it is usually observed at high temperature and high electric field. For the Poole-Frenkel emission, a plot of $\ln(J/E)$ versus $E^{1/2}$ is linear and the slope of the plot gives the Poole-Frenkel slope. From the plot in Fig. S9(a), Poole-Frenkel slope between 4e-5 and 4.5e-5 eVV$^{-1/2}$m$^{1/2}$ was obtained for reverse voltage higher than 30 V. Trap barrier height of 0.38 eV was obtained from the Arrhenius plot shown in Fig. S9(b).

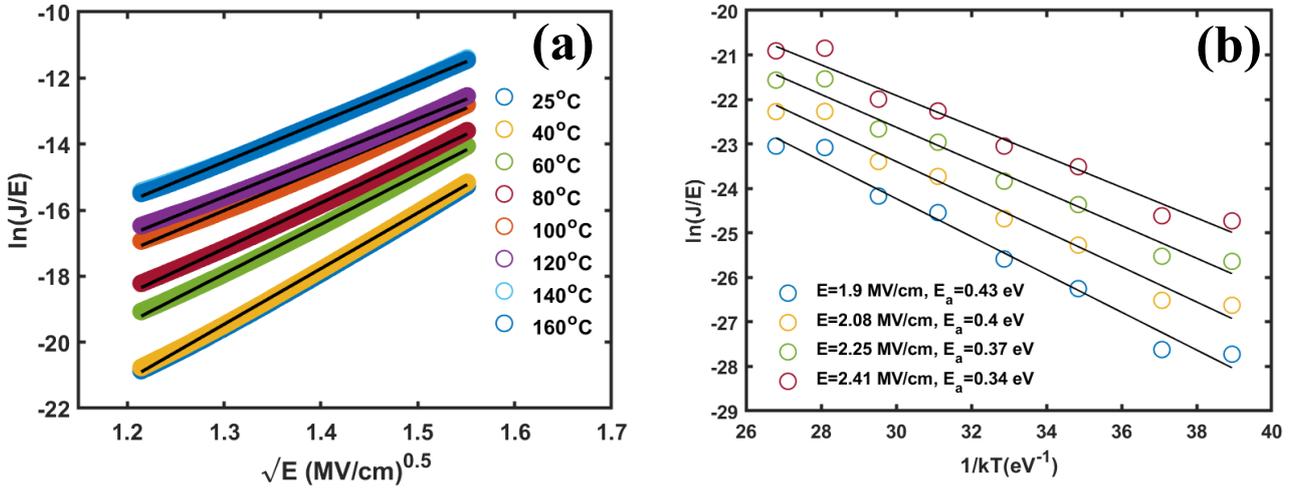

FIG. S9. (a) Characteristic plot of Poole-Frenkel conduction and (b) Arrhenius plot for the MIS diode under high reverse bias between 25 °C and 160 °C.

# X. Temperature dependent reverse bias *J-V* fits for the MIS diodes

Figs. 5(b) and 5(c) in the manuscript showed the reverse bias *J-V* fits for the MIS diode at 40 °C and 120 °C respectively. The fits at the other temperatures are shown in Fig. S10.

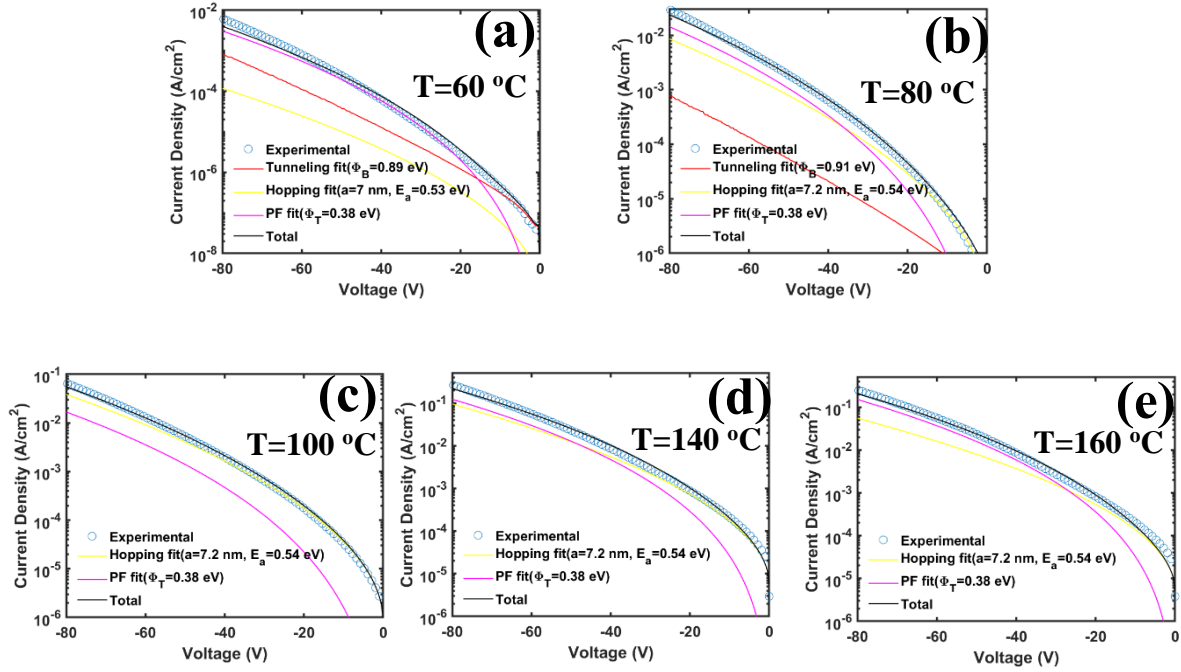

FIG. S10. Temperature dependent reverse bias *J-V* characteristics for the MIS diode and their corresponding fits at (a) 60 °C, (b) 80 °C, (c) 100 °C, (d) 140 °C and (e) 160 °C.